\documentclass{ws-mpla}
\usepackage{subfigure}
\usepackage{hyperref}

\begin{document}

\markboth{Gustaaf Brooijmans}
{After the Standard Model: New Resonances at the LHC}

\catchline{}{}{}{}{}

\title{AFTER THE STANDARD MODEL: NEW RESONANCES AT THE LHC}

\author{\footnotesize GUSTAAF BROOIJMANS}

\address{Physics Department, Columbia University\\
New York, New York 10027,
USA\\
gusbroo@nevis.columbia.edu}

\maketitle

\pub{Received (Day Month Year)}{Revised (Day Month Year)}

\begin{abstract}
Experiments will soon start taking data at CERN's Large Hadron Collider (LHC)
with high expectations for discovery of new physics phenomena.  Indeed,
the LHC's unprecedented center-of-mass energy will allow the experiments to
probe an energy regime where the standard model is known to break down.
In this article, the experiments' capability to observe new resonances in 
various channels is reviewed.
\end{abstract}

\ccode{PACS Nos.: 12.60.-i, 12.60.Cn, 12.90.+b.}

\section{Introduction}	
The standard model of particle physics has proven to be a remarkably accurate
framework for the description of the interactions of particles at energies up 
to almost 1~TeV~\cite{Alcaraz:2007ri,:2005ema}.  At 1.7~TeV, however, 
the standard model longitudinal $W$ boson scattering cross-section violates
the unitarity bound~\cite{Lee:1977eg}.  
One solution to this problem is the introduction of the so-called Higgs 
boson~\cite{Higgs:1964ia,Higgs:1964pj,Higgs:1966ev,Englert:1964et,Guralnik:1964eu},
which in the standard model can also generate the fermion masses.  
The latter allows the decoupling of the mechanism responsible
for fermion masses from the standard model interactions.
However, quadratically divergent radiative corrections suggest the 
standard model Higgs boson mass 
is close to the limit of validity of the theory, and experimental
constraints~\cite{Alcaraz:2007ri} imply its mass is less than approximately 200~GeV.
This suggests the scale of new physics is at or below 1~TeV, although in principle, 
if one accepts a high level of fine-tuning, 
with the addition of a $m_H = 150$ GeV Higgs boson the list
of existing particles could be ``complete'', in analogy with Mendeleev's table 
in chemistry.  No new physics would then appear below the Planck scale of
$10^{19}$ GeV.

In addition to these ``technical'' issues in the standard model, it is good to
re-emphasize that it is a theory of interactions, in which the properties of 
the interaction bosons in terms of couplings (to fermions and each other), 
propagation and masses are 
linked and testable, but the properties of fermions are inputs.  The standard 
model does not give any information regarding the nature of particles and 
leaves many fundamental questions unanswered.  For example: What is color?  
Or electric charge?  Why are they quantized?  Are these dynamic or static
properties of particles?  Are there only three generations?  Why are there
no neutral, colored fermions\footnote{A recent paper~\cite{Shrock:2008sb} explores the consequences
of modifying hypercharge to allow for such particles.}?  
Is there a link between particle and nucleon masses?

There are many models of physics beyond the standard model that address at 
least the technical issues,
including supersymmetry~\cite{Martin:1997ns}, 
little Higgs~\cite{ArkaniHamed:2001nc}, and models with additional space
dimensions~\cite{ArkaniHamed:1998rs,Randall:1999ee}.  Most of these predict 
the existence of new particles that should be produced and detected at the 
LHC.  Hopefully, the observation of such new particles and the measurement of 
their properties will also help us understand the patterns observed in 
the properties of the standard model fermions.

Looking at the three currently known generations of fermions, it appears that
within a generation\footnote{This is if the tau lepton and tau neutrino
are part of the same generation as the top and bottom quarks, and similarly
for muon, muon neutrino, and charm and strange quarks, as is usually assumed.},
the more a fermion interacts, the larger its mass:
colored particles are heavier than their color-neutral counterparts;
up-type quarks, with electric charge $|q|=2/3$ are heavier than bottom-type
quarks of charge $|q|=1/3$; charged leptons are heavier than neutral leptons.
This pattern suggests that the fermion masses might be related to a more 
complex mechanism leading to an indirect link between masses and standard
model interactions.  In that case, the Higgs boson might only be relevant
to the unitarization of massive vector boson scattering, which would relax
the existing constraints on its mass.

\section{Signals of Parity Restoration}

Maximal violation of parity in weak interactions in ``catastrophic'' since
for massive fermions helicity depends on the reference frame.
Therefore, a deeper 
understanding of the nature of fermion spin, and thus the weak 
interaction would allow progress similar to the discovery of a Higgs
boson: the mechanism of restoration of parity symmetry might lead to 
an understanding of fermion masses.

The primary signals of parity restoration would be the presence of a
$W$-like boson coupling to right-handed fermions with weak coupling strength.  
As in the standard model,
the neutral member of such a triplet could mix with the other neutral 
electroweak bosons and 
have a different mass than its charged companions.  We generically 
denote these extra bosons as $W'$ and $Z'$.  At energies much larger than 
$m_{W'},m_{Z'}$, the left- and right-handed interactions have the same 
strength and the symmetry is restored.

$W'$ and $Z'$ bosons are predicted to exist not just in
left-right symmetric models, but also many constructions inspired by grand unification.
The production cross-sections and decay branching ratios (and widths) 
depend on the specific model
through the couplings~\cite{Rizzo:2006nw}, but in most cases the $Z'$ boson
decay to a pair of leptons remains the ``golden'' signature.

\subsection{$Z'$ Boson Decays To Leptons}

\begin{figure}[ph]
\centerline{
\subfigure {
\psfig{file=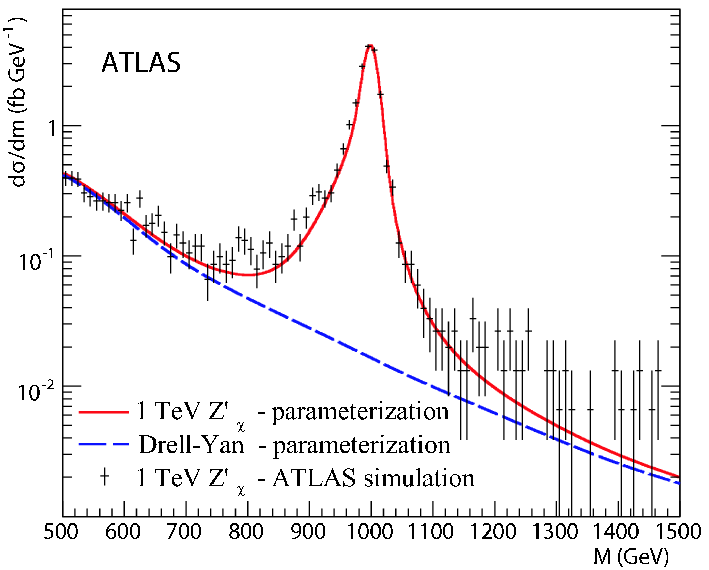,width=0.48\textwidth}
}
\subfigure {
\psfig{file=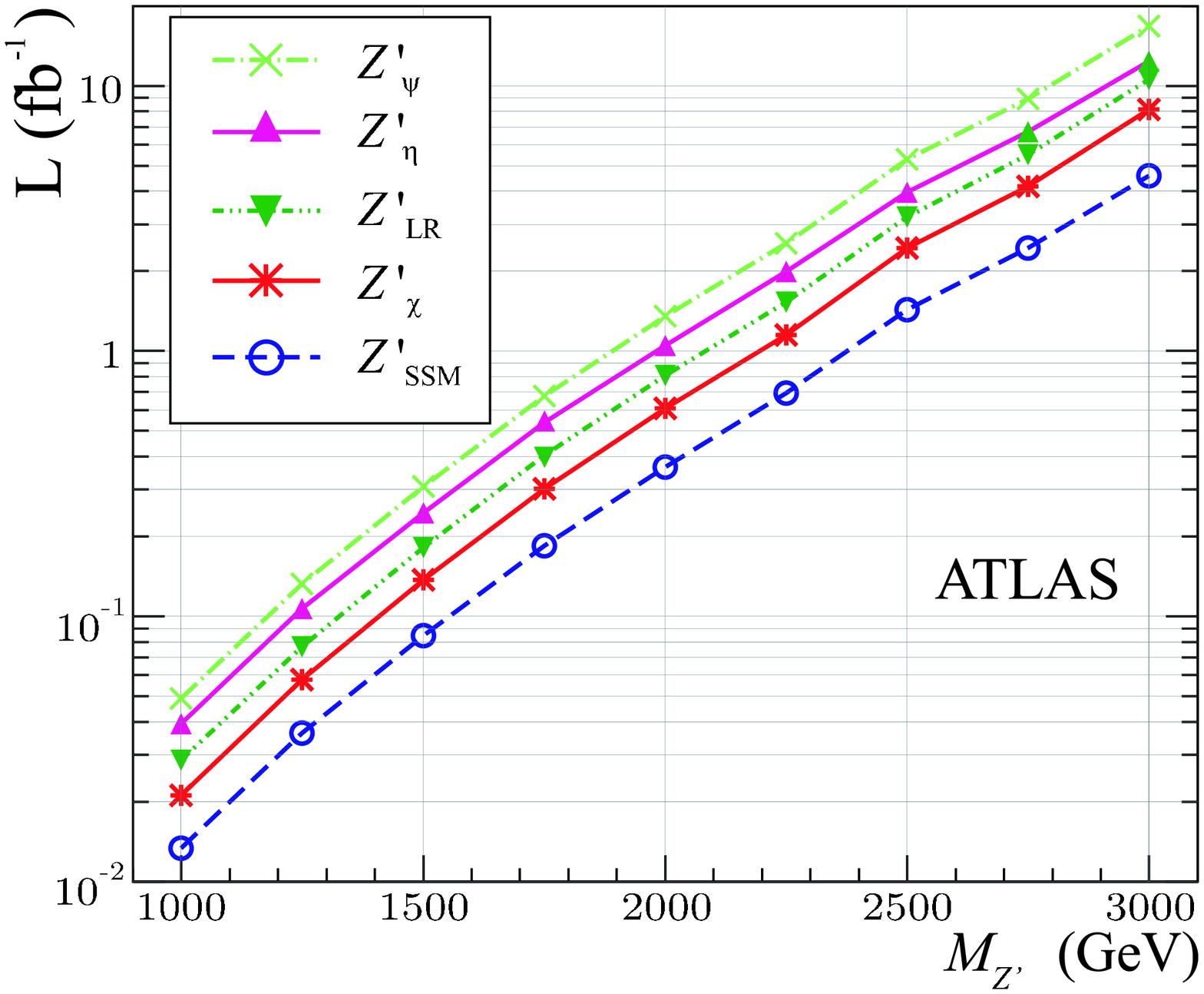,width=0.48\textwidth}
}}
\vspace*{8pt}
\caption{Left: mass spectrum for a $m_{Z'}=$ 1 TeV $Z^{\prime}_{\chi}\to e^+e^-$ signal obtained with 
ATLAS full simulation (histogram) and a parameterization (solid line) used to estimate 
the signal significance.  Right: Integrated luminosity needed for 5$\sigma$ discovery 
of a $Z^{\prime}\to e^+e^-$ signal in various models as a function of the $Z^{\prime}$ boson mass.
\protect\label{fig:mll}}
\end{figure}

The most promising decay channels are $Z' \to e^+e^-$ and $Z' \to \mu^+\mu^-$,
where the only irreducible background comes from the Drell-Yan continuum, and
instrumental backgrounds from misidentification of jets as leptons are 
typically significantly smaller.  Current limits from searches for such objects at the 
Fermilab Tevatron imply masses larger than approximately 1 TeV for couplings
identical to the standard model $Z$ boson~\cite{cdfee08,cdfmumu08}.
Figure~\ref{fig:mll} (left) shows the production 
cross-section times branching ratio folded with detector efficiency and resolution 
for a $m_{Z'} =$ 1 TeV E$_6$-inspired 
$Z'_\chi$ boson as a function of the invariant 
mass of the reconstructed electron-positron pair, as obtained in a full simulation
of the ATLAS detector~\cite{atlascsc}.  The luminosity needed to discover such a 
resonance is shown as a function of the resonance mass in Fig.~\ref{fig:mll} (right)
for a few models.  Tens of inverse picobarns of data at a center-of-mass energy of 
14 TeV suffice to extend the reach beyond the Fermilab Tevatron limits, and with 
10 fb$^{-1}$ a $m_{Z'}=$ 3 TeV resonance will be observable.  Thanks to the excellent
performance of the detectors, the sensitivity is similar in the muon 
channel~\cite{Ball:2007zza}.  The ultimate LHC sensitivity is expected to be in the 
range of 5--6 TeV for most models.

One difficulty with these searches originates in the a priori unknown mass of the 
resonance.  The data analysis therefore often uses a moving 
``window'' in the invariant mass distribution
to compare the data with the expectation from standard model backgrounds, either 
by counting events or comparing a signal plus background distribution to a background-only
one.  When using such a strategy, it is important to factor in the number of mass
windows in which the search is conducted, as was done by CDF~\cite{:2007sb}.  An 
alternative is to use a maximum likelihood fit where masses, cross-sections 
and widths are free parameters.  Toy simulations show that 
the real sensitivity is typically about 20\% lower
compared to cases where the ``look elsewhere'' effect is ignored~\cite{atlascsc}.

Once a signal is established, its nature can be determined by measuring not only
its production cross-section and width (if larger than the detector resolution), but also 
its spin and couplings to fermions.  By measuring the angle between one of the leptons
and the beam direction, the spin of a Randall-Sundrum graviton can be determined with
90\% confidence in 
100 fb$^{-1}$ for a resonance mass up to 1720 GeV~\cite{Allanach:2000nr}.
In case of a vector resonance, the couplings can be determined by 
measuring the forward-backward asymmetry.  
For a mass of 1 (3) TeV, an integrated luminosity of 10 (400) fb$^{-1}$
allows the distinction between various E$_6$ model points or a left-right symmetric
signal with a significance larger than 3$\sigma$~\cite{Cousins:916380}.

\subsection{Other $Z'$ Boson Decays}

$Z'$ boson decays to light quarks are significantly more difficult to detect given 
the much larger backgrounds from QCD dijet production and an energy resolution which 
is significantly worse for jets than leptons.  Such signals are observable however,
but with a sensitivity which is typically one or two orders of magnitude worse than 
in the dilepton channel~\cite{Henriques:682130}.  Decays to pairs of top quarks 
require specialized techniques and will be discussed later.

Another possibility in left-right symmetric models is that the $Z'$ boson decays to
right-handed neutrinos, provided these are light enough.  The right-handed neutrino can 
then decay to a lepton and two jets: $N_R \to \ell W_R^*$, $W_R^* \to 
q \bar{q'}$.  In this case the final state contains two leptons and four jets, 
and since the right-handed neutrino is a 
Majorana particle, the final state leptons can have the same sign.  If furthermore
the right-handed neutrino is relatively light, it will be highly boosted and the 
collimation of the lepton and the jets from its decay will lead to the lepton 
being embedded in the jets.  This decay chain has been studied and, depending 
on the mass of the heavy neutrino, discovery can be achieved for $Z'$ boson masses
up to 5 TeV in 300 fb$^{-1}$ of data~\cite{Ferrari:684146}.
Other decay chains are possible: if there is a mass hierarchy between different 
flavors of right-handed neutrinos for example, additional leptons will probably
be present.  

\subsection{$W'$ Boson Production and Decay}

\begin{figure}[ph]
\centerline{
\psfig{file=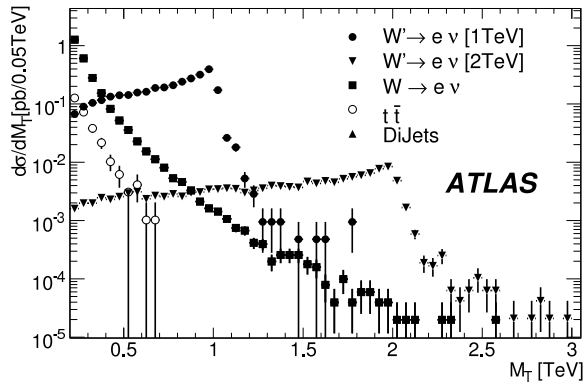,width=0.6\textwidth}
}
\vspace*{8pt}
\caption{Transverse mass spectrum in $W' \to e \nu_{(R)}$ decays in a full 
simulation of ATLAS data after requiring that the reconstructed 
event contain a single isolated lepton with $p_T >$ 50 GeV, missing transverse
energy $>$ 50 GeV, and a ``lepton fraction'' 
$\sum p_T^{leptons}/(\sum p_T^{leptons}+ \sum E_T) >$ 0.5.
\protect\label{fig:mt}}
\end{figure}

The total $W'$ boson production rate at the LHC is not very dependent on the boson's 
helicity couplings to fermions, 
however the interference with the standard model $W$ boson, which shapes  
$d\sigma/dM$, is key to the determination
of these couplings~\cite{Rizzo:2007xs} and has so far rarely been included in experimental 
studies.  A $W'$ boson can be searched for in the very clean $W' \to \ell \nu$
channel by studying the transverse mass distribution in events with a single isolated 
lepton and missing transverse energy.  If such a decay mode is open, resonances of mass
up to approximately 3 (4.5) TeV can be discovered with as little as 1 (10) fb$^{-1}$ of 
data~\cite{atlascsc,Ball:2007zza}.  Transverse mass spectra for a few signal points and 
dominant backgrounds are shown in Fig.~\ref{fig:mt}.  
Even though the resolution function is very different between
muons and electrons, because of very low backgrounds 
the sensitivity is very similar in both channels.
However, in a purely left-right model for example, this decay is forbidden.

\begin{figure}[htbp]
\begin{center}
\subfigure[]{
\includegraphics[width= 
0.45\textwidth]{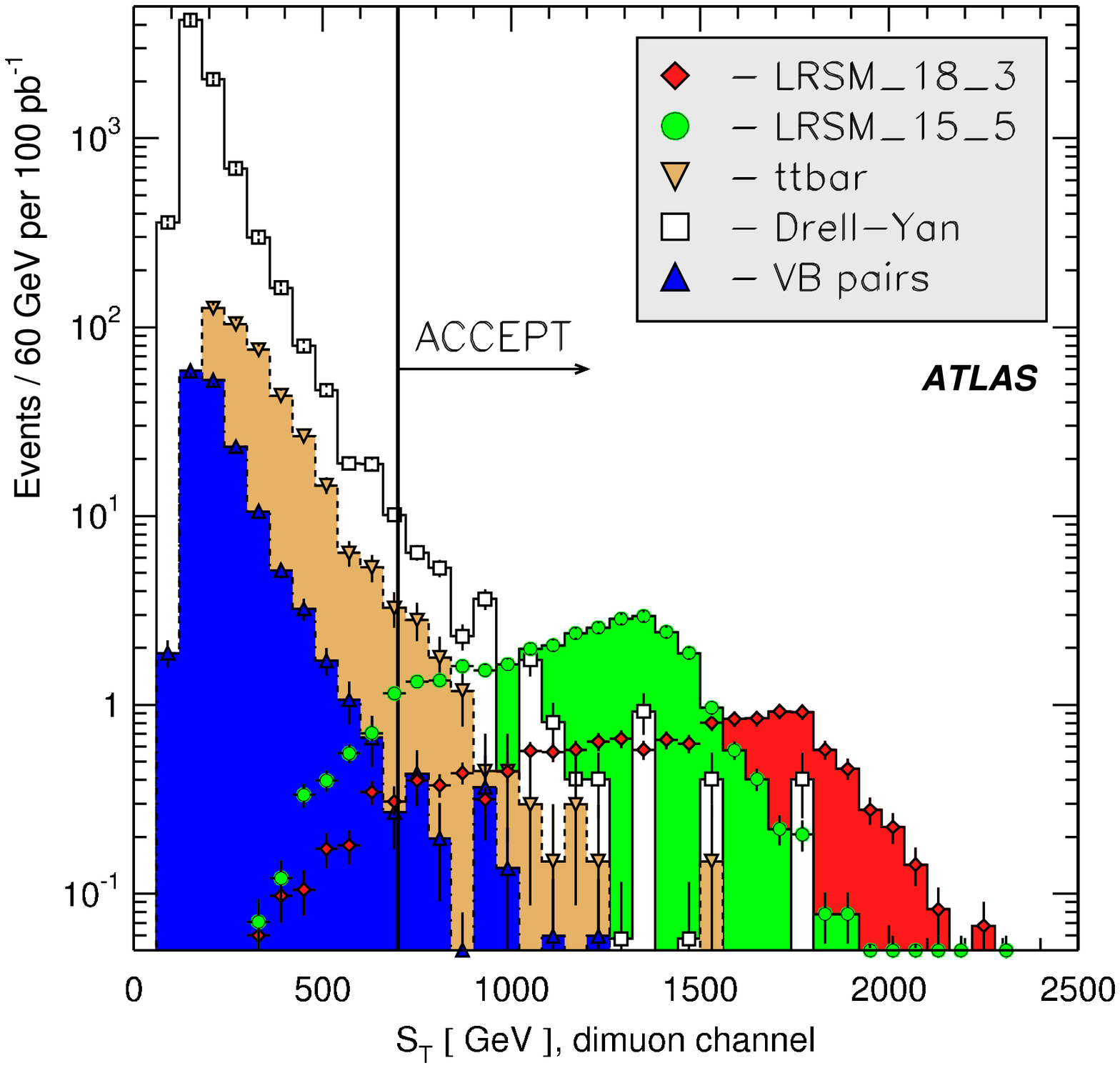}
}
\subfigure[]{
\includegraphics[width= 
0.45\textwidth]{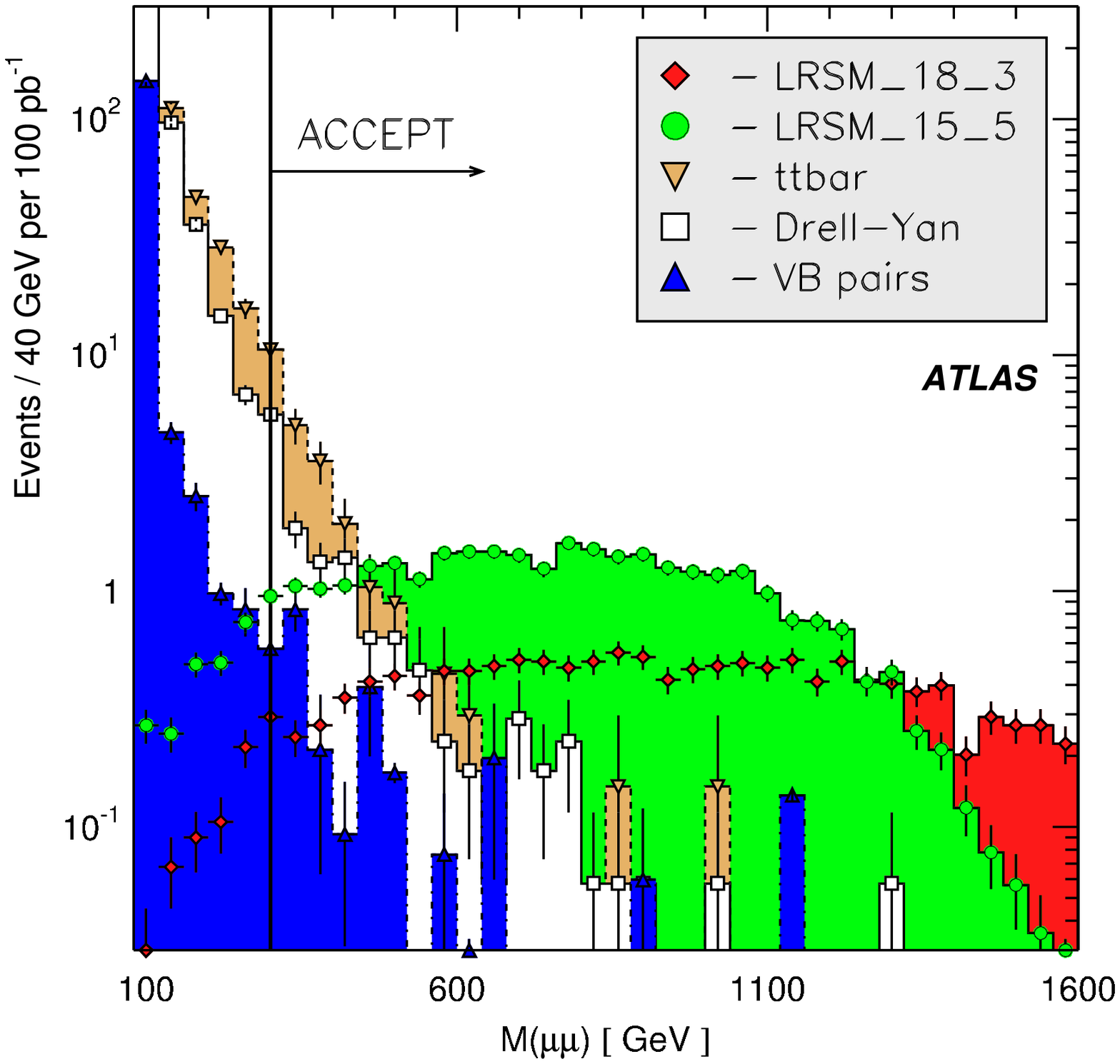}
}
\caption{Sum of the transverse energies of the two muons and two leading jets ($S_T$) (a), 
and dilepton invariant mass (b) for backgrounds and two signal points in the search
for $W'$ bosons decaying following the chain 
$W' \to \mu N_R, N_R \to \mu W'^*, W'^* \to q \bar{q'}$.  The signal points 
LRSM\_18\_3 and LRSM\_15\_5 correspond to masses $m_{W'} =$ 1800 GeV, $m_{N_R} =$
300 GeV and $m_{W'} =$ 1500 GeV, $m_{N_R} =$ 500 GeV respectively.
\label{fig:lrsm1}}
\end{center}
\end{figure}

\begin{figure}[htbp]
\begin{center}
\subfigure[]{
\includegraphics[width= 
0.45\textwidth]{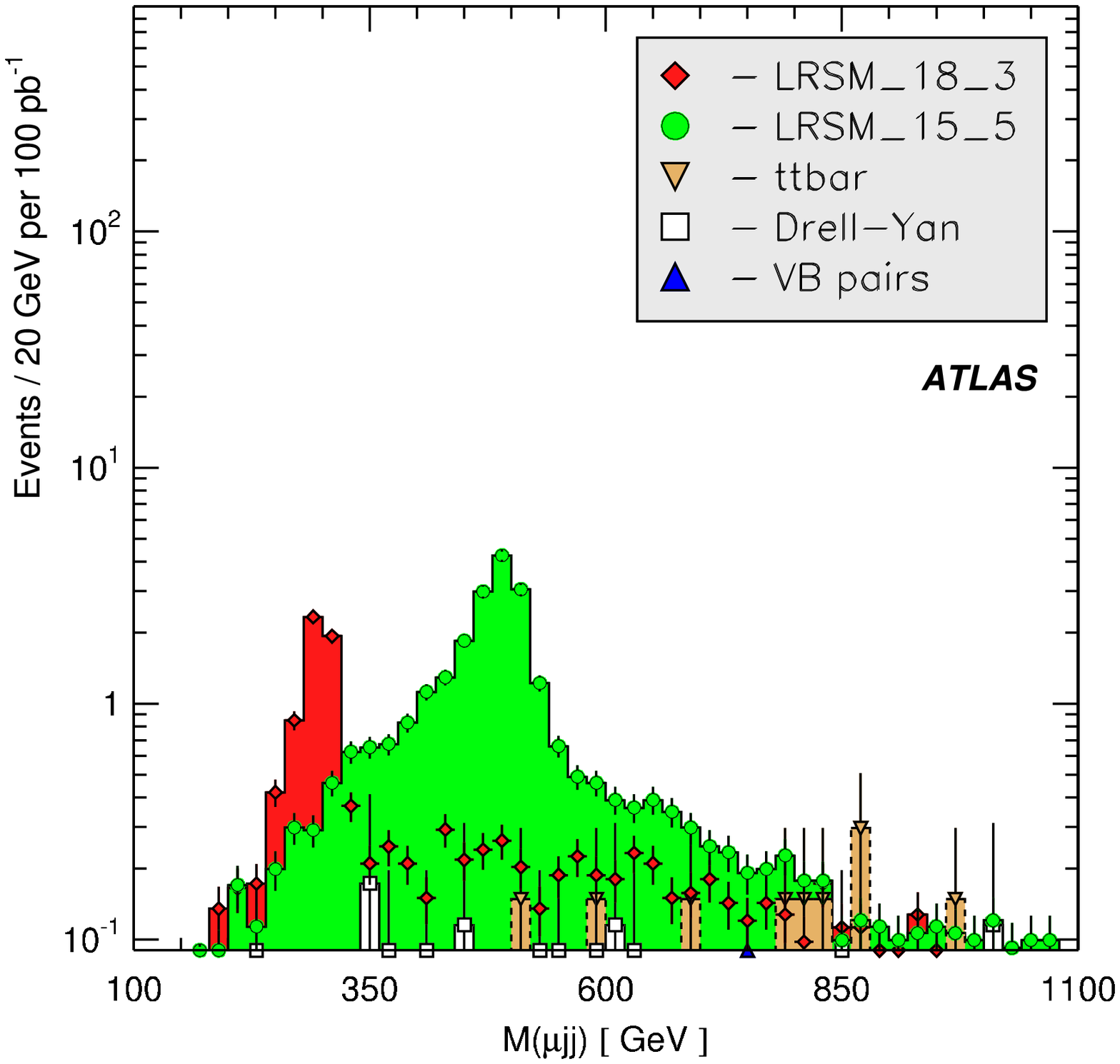}
}
\subfigure[]{
\includegraphics[width= 
0.45\textwidth]{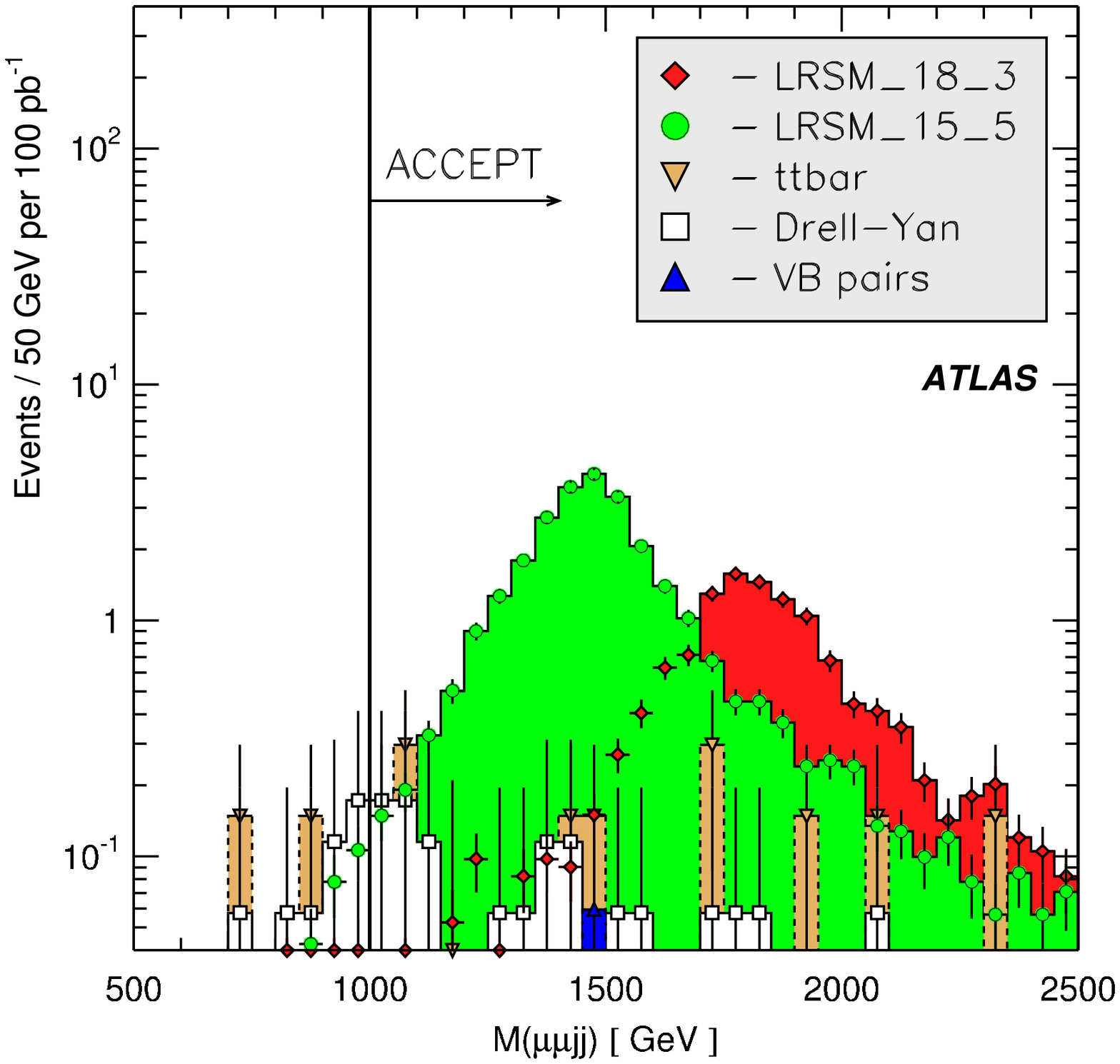}
}
\caption{Reconstructed candidate (a) $N_R$ neutrino and (b) $W'$ boson masses after 
cuts on $S_T$ and the dilepton mass.  The signal points 
LRSM\_18\_3 and LRSM\_15\_5 correspond to masses $m_{W'} =$ 1800 GeV, $m_{N_R} =$
300 GeV and $m_{W'} =$ 1500 GeV, $m_{N_R} =$ 500 GeV respectively.
\label{fig:lrsm2}}
\end{center}
\end{figure}

If the right-handed neutrino ($N_R$) mass is smaller than that of the $W'$ boson, the 
$W' \to \ell N_R$ channel opens up. Of course, if the $N_R$ neutrino is 
stable on the scale of the 
detector, the signature from $W' \to \ell N_R$ is very similar to the classical one.
The decay chain $N_R \to \ell W'^*$, $W'^* \to q \bar{q'}$ may also be observable.
The final state contains at least two hard leptons and two hard jets.
Note that depending on the mass differences between the $W'$ and the heavy 
neutrino $N_R$, one of the leptons could be very close to one of the jets.  The dominant
standard model backgrounds to these searches are $t\bar{t}$ production, production
of $Z/\gamma^*$ in association with hard jets, and diboson production.  These are
effectively suppressed by cuts on the summed transverse energy of the two leptons
and two jets ($S_T$), and the invariant mass of the two leptons.  This is illustrated
in Fig.~\ref{fig:lrsm1} for both signal and backgrounds from a study of the dimuon channel 
based on a full simulation
of the ATLAS detector~\cite{atlascsc}.
After these cuts, the candidate $N_R$ neutrino mass can be reconstructed by taking 
the invariant mass of the two jets with each of the leptons separately.  In the analysis shown
here, the combination with the smallest invariant mass is kept as the $N_R$ candidate.
The other lepton can then be added to form the $W'$ boson.  Both invariant mass 
distributions are shown in Fig.~\ref{fig:lrsm2} for the muon channel.
With this analysis, the signal point with $m_{W'} =$~1500 GeV, $m_{N_R} =$~500 GeV can
be discovered with as little as 20~pb$^{-1}$ of data, while the 
$m_{W'} =$~1800 GeV, $m_{N_R} =$~300 GeV point requires almost 200~pb$^{-1}$.  Since
these results were obtained by counting events above background after cuts, a further 
increase in the sensitivity should be attainable with a more sophisticated 
analysis technique based on shapes of distributions.

\begin{figure}[hbtp]
\centerline{
\psfig{file=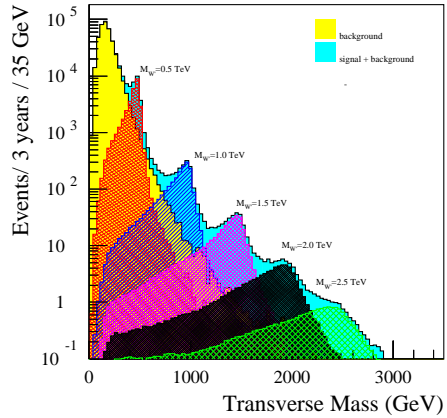,width=0.48\textwidth}
}
\vspace*{8pt}
\caption{Transverse mass spectrum for $W' \to WZ \to \ell\nu \ell\ell$ decays for
various signal masses and the standard model background in an ATLAS simulation 
of 300 fb$^{-1}$ of data.
\protect\label{fig:mt2}}
\end{figure}

Another promising channel is 
the decay chain $W' \to WZ \to \ell\nu \ell\ell$, where again the transverse mass 
distribution, but now built with three charged leptons, is key in the analysis.  This
distribution is shown for various $W'$ boson masses and for the standard model 
backgrounds for 300 fb$^{-1}$ of ATLAS data in Fig.~\ref{fig:mt2}.  
In a dataset that 
size, a $W'$ boson with couplings identical to the standard model $W$ boson is 
observable for masses up to 2.8 TeV~\cite{Arik:684195}.  If after the $W' \to WZ$
decay one allows one of the $W$ or $Z$ bosons to decay hadronically, the overall 
branching ratio goes up along with the backgrounds.  Since the $W$ or $Z$ boson
is highly boosted, the quarks from the decay are collimated and form a single 
hadronic jet.  This jet can however be distinguished from a light quark jet through
its mass splitting scale, which will be explained in detail below.  The 
backgrounds from $W/Z$+jet production are not well known, and no reliable estimates of 
sensitivity exist at this time.

Identification of the $W'^+ \to t\bar{b}$ decay (and its charge conjugate)
requires high-$p_T$ $b$- and top-tagging, the specifics of which will 
be discussed later.  Early studies show that $W'$ bosons from little Higgs
models can be discovered in this channel for masses up to 2.5 TeV in as little as
30 fb$^{-1}$ of data~\cite{GonzálezdelaHoz:814346}.  

\subsection{Exotic Quarks}

In most cases, the existence of $Z'$ bosons requires the existence of new exotic
quarks or leptons~\cite{Langacker:2008yv}.  Such quarks could be pair-produced and
decay to a standard model gauge boson and quark.  For down-type quarks one would for 
example have $pp \to D\bar{D}$, with $D \to Wu$ or $D \to Zd$.  The sensitivity of 
the ATLAS experiment to such quarks in final states with at least two leptons has been
investigated~\cite{Mehdiyev:2006tz,Mehdiyev:2007pf}, showing that these quarks can 
be discovered up to masses of about 1 TeV with 100 fb$^{-1}$ of data.
\begin{figure}[htbp]
\begin{center}
\subfigure[]{
\includegraphics[width= 
0.45\textwidth]{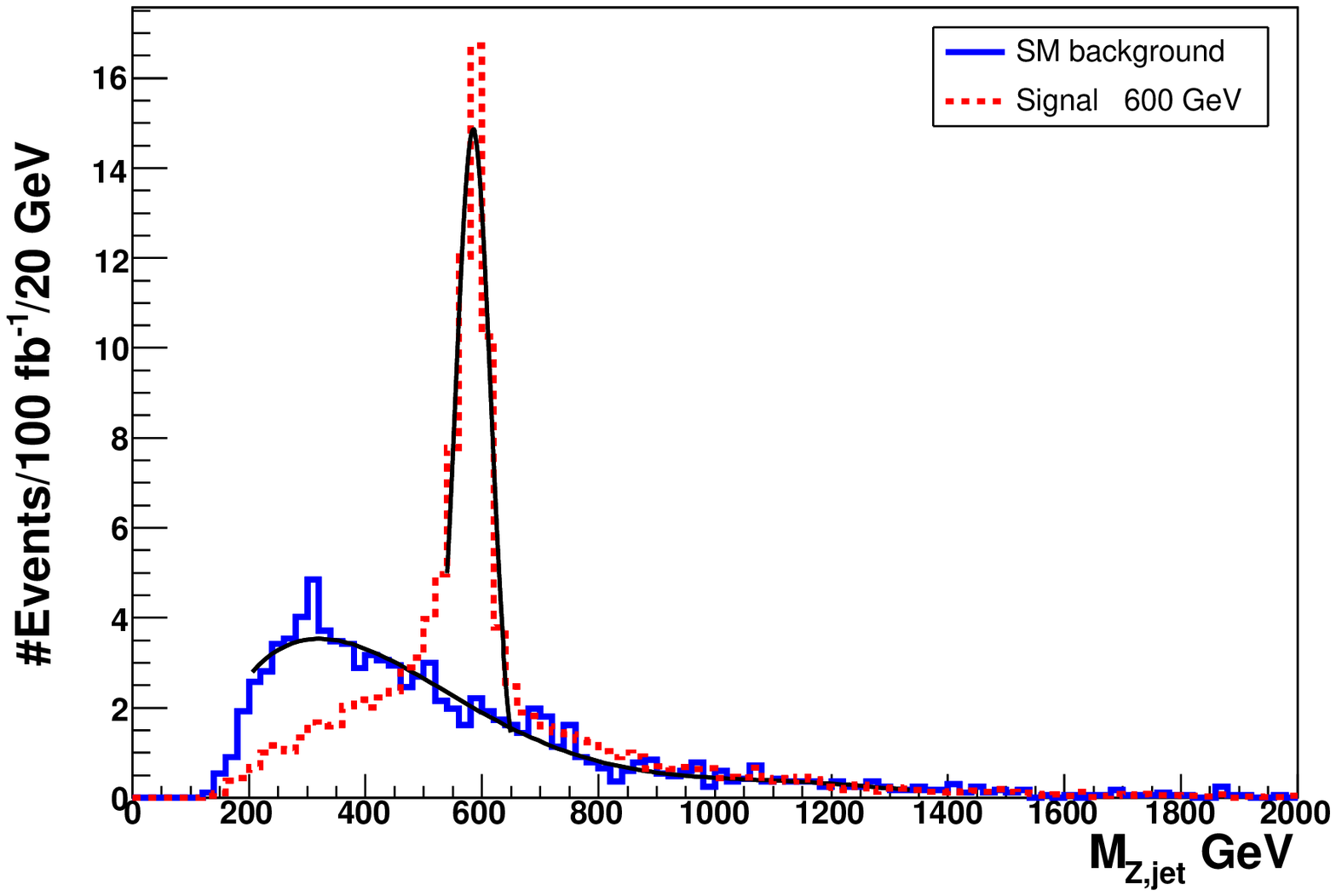}
}
\subfigure[]{
\includegraphics[width= 
0.45\textwidth]{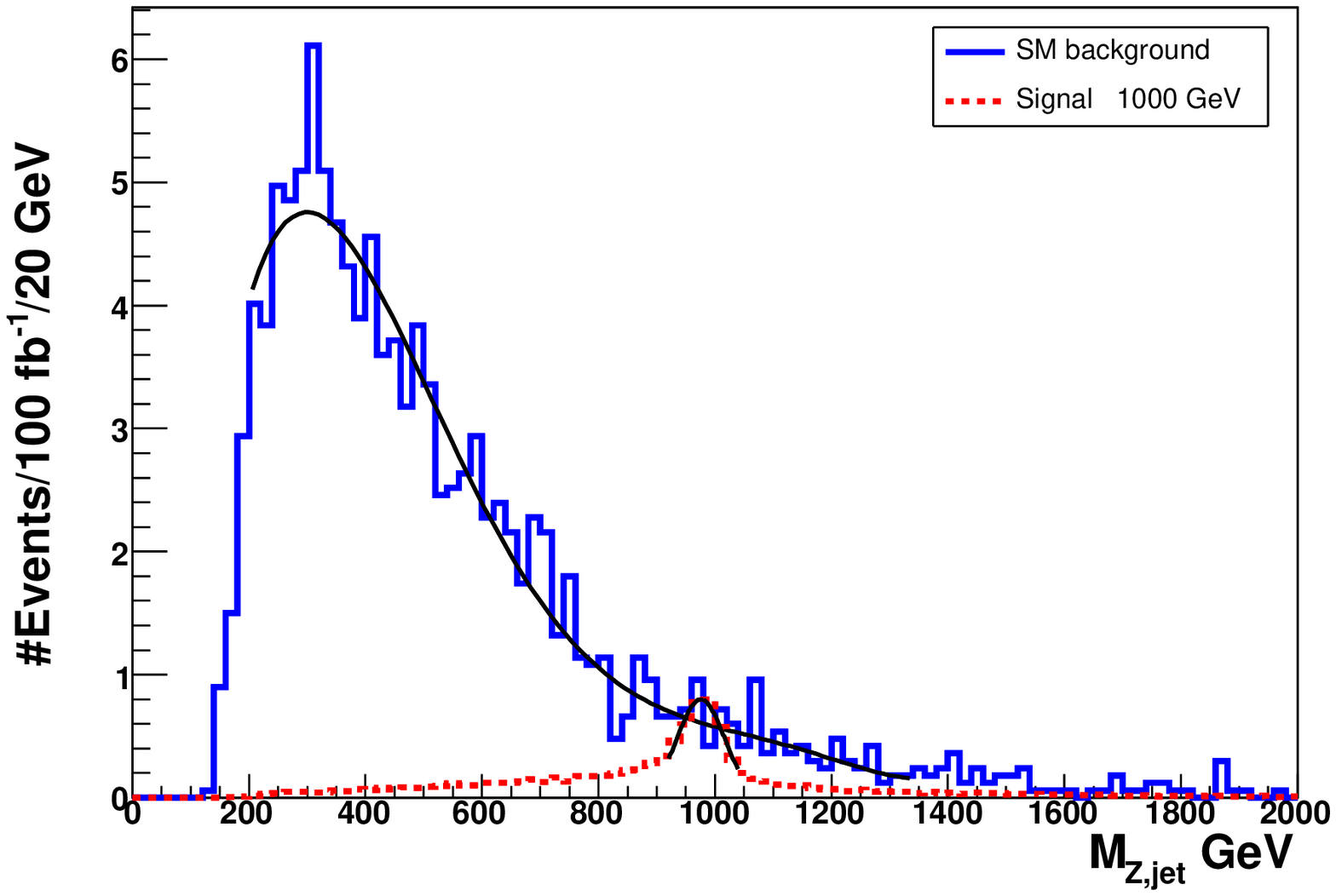}
}
\caption{Invariant $Z$ boson plus jet mass spectra for both signal and background 
in the search for the exotic quark 
decay $D \to Zd$ in the $D\bar{D} \to Z d Z\bar{d} \to \ell \ell j \nu \nu j$ channel 
for two different $D$ quark masses: (a) $m_D=$ 600, (b) $m_D=$ 1000 GeV.
\label{fig:exoq}}
\end{center}
\end{figure}
Figure~\ref{fig:exoq} shows the $Z$ boson plus jet invariant mass spectra that are obtained
by combining the leptonically decaying $Z$ boson with each of the hard jets in events with 
two isolated leptons, two hard jets and substantial missing transverse energy, as expected
from the $D\bar{D} \to Z d Z\bar{d} \to \ell \ell j \nu \nu j$ decay chain.

\section{Extra Dimensions}

A promising approach to quantum gravity consists in supposing the existence of 
extra space dimensions, a 
strategy known as string theory.  To explain that these additional space dimensions 
are not observed, string theorists hypothesize that 
they are compactified, i.e. folded on themselves at a length scale much smaller than what
is accessible in experiments.  This compactification scale is typically assumed to be 
the scale of gravity, i.e. 10$^{19}$ GeV.  However, in the late 90's people realized
this scale may actually be much lower, in reach of current or near-future experiments.

\subsection{``ADD'' Extra Dimensions}

In the so-called ``ADD'' model, after Arkani-Hamed, Dimopoulos and 
Dvali~\cite{ArkaniHamed:1998rs}, standard model fields are confined to a
3 + 1 dimensional subspace (``brane'') and gravity is allowed to propagate
in the ``bulk'' space of extra dimensions.  The extra dimensions have a
flat metric, and the reason gravity appears very weak to us is that it is
only felt when the graviton goes through the standard model brane.

The edges of the compactified extra dimensions are identified, leading to boundary
conditions on the wave functions of particles propagating along these directions.
Momentum along the extra dimensions is therefore quantized, and appears as mass
to observers in the standard model brane.  Gravitons thus
acquire a ``mass'' proportional to their extra-dimensional momentum.
In the ADD model, the mass splittings between these different states, commonly called 
``excitations'', are small, and the ``tower''
of graviton excitations appears as a continuous distribution in mass.  While the coupling
to a single graviton remains extremely weak, there are a very large number of accessible
states leading to a very large phase space, and observable cross-sections.  Furthermore,
since the graviton couples to energy-momentum, all processes are affected.

At colliders, searches for evidence for the presence of such extra dimension have been
conducted in two categories of events: those where a graviton is directly produced
and immediately disappears into the bulk, and those where gravitons interfere
with a standard model process.  At hadron colliders, the former lead to signatures 
with a single hard jet or gauge boson accompanied by a large amount of missing 
transverse energy due to the escaping graviton.  The predicted signal consists of
an excess of events at high transverse momentum.   Searches at the Fermilab Tevatron 
result in lower limits on the compactification scale of the order of 1 
TeV~\cite{Abazov:2008kp,Aaltonen:2008hh}.  The most sensitive process to study in 
the second category appears to be Drell-Yan production of charged lepton pairs.  In addition
to the high mass tail, the angular distribution of the leptons can be used to 
increase the sensitivity to the presence of a spin-2 particle.  Based on this
technique, the D\O\ experiment
has set limits~\cite{Abazov:2008as} on the compactification scale between 1 and 2 TeV 
depending on the number of extra dimensions.

\subsection{Warped Extra Dimensions}

\begin{figure}[hbtp]
\centerline{
\psfig{file=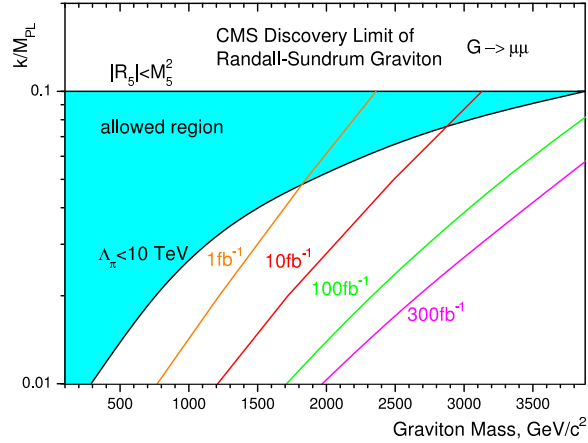,width=0.60\textwidth}
}
\vspace*{8pt}
\caption{Reach of the CMS experiment as a function of the coupling parameter 
$c= k/M_{Pl}$ 
and the graviton mass for various values of integrated luminosity. 
The left side of each curve is the region where significance exceeds 5$\sigma$,
and the blue shaded region is favored in terms of addressing the hierarchy
between the Planck and electroweak scales.
\protect\label{fig:grav1}}
\end{figure}

Randall and Sundrum proposed~\cite{Randall:1999ee} 
a model in which the hierarchy between the electroweak
and Planck scale is generated by a warped metric in an extra dimension.
In this model there are two branes, with the standard model fields
confined to one, and the metric between the two warped by 
$e^{-2kr_c\phi}$, where $k$ is the warp factor, $r_c$ the compactification
radius, and $\phi$ the coordinate along the extra dimension.  With $kr_c \approx 50$,
a hierarchy of $10^{15}$ is created between the branes located at $\phi = 0$ and $\pi$,
allowing the generation of TeV-scale masses from the 10$^{19}$ GeV Planck scale.
In this scenario, instead of an almost continuous tower, a 
small number of heavy graviton excitations $G$ exist, but these couple with electroweak 
strength and are therefore individually observable.  They are widely spaced 
resonances decaying mainly to pairs of standard model particles.  Searches 
at the Tevatron have set limits between a few hundred GeV and one TeV on 
the mass of such excitations, depending on the magnitude of the warp factor 
$k$~\cite{Abazov:2007ra,:2007sb}.  The LHC experiments should be able to 
cover the entire region of interest (i.e. where this solution generates the 
hierarchy between the Planck and electroweak scales), as shown in a CMS 
study~\cite{Ball:2007zza} of $G \to \mu^+ \mu^-$ in Fig.~\ref{fig:grav1}.

An interesting variation on this model has the standard model fermions located
along the extra dimension~\cite{Davoudiasl:2000wi,Agashe:2003zs}.  
The fermion masses and mixings are then generated 
by geometry, and the heavier fermions as well as gauge boson excitations are located
close to the ``TeV'' brane.  The most promising channels for discovery then
become the direct production of gauge boson excitations, and in particular
the first excitation of the gluon.  This is expected to have a mass larger
than about 2~TeV based on constraints from precision electroweak 
measurements~\cite{Davoudiasl:2000wi}.
However, even though the object is strongly interacting, the production cross-section is 
relatively small~\cite{Lillie:2007yh} because of the small overlap of its 
wavefunction in the extra
dimension with the light fermions' wavefunctions.  Correspondingly, the dominant branching ratio
is to pairs of top quarks or longitudinal $W$ or $Z$ bosons, and the width 
can be large~\cite{Djouadi:2007eg}.  

\begin{figure}[htbp]
\begin{center}
\subfigure[]{
\label{fig:dr1}
\includegraphics[width= 
0.45\textwidth]{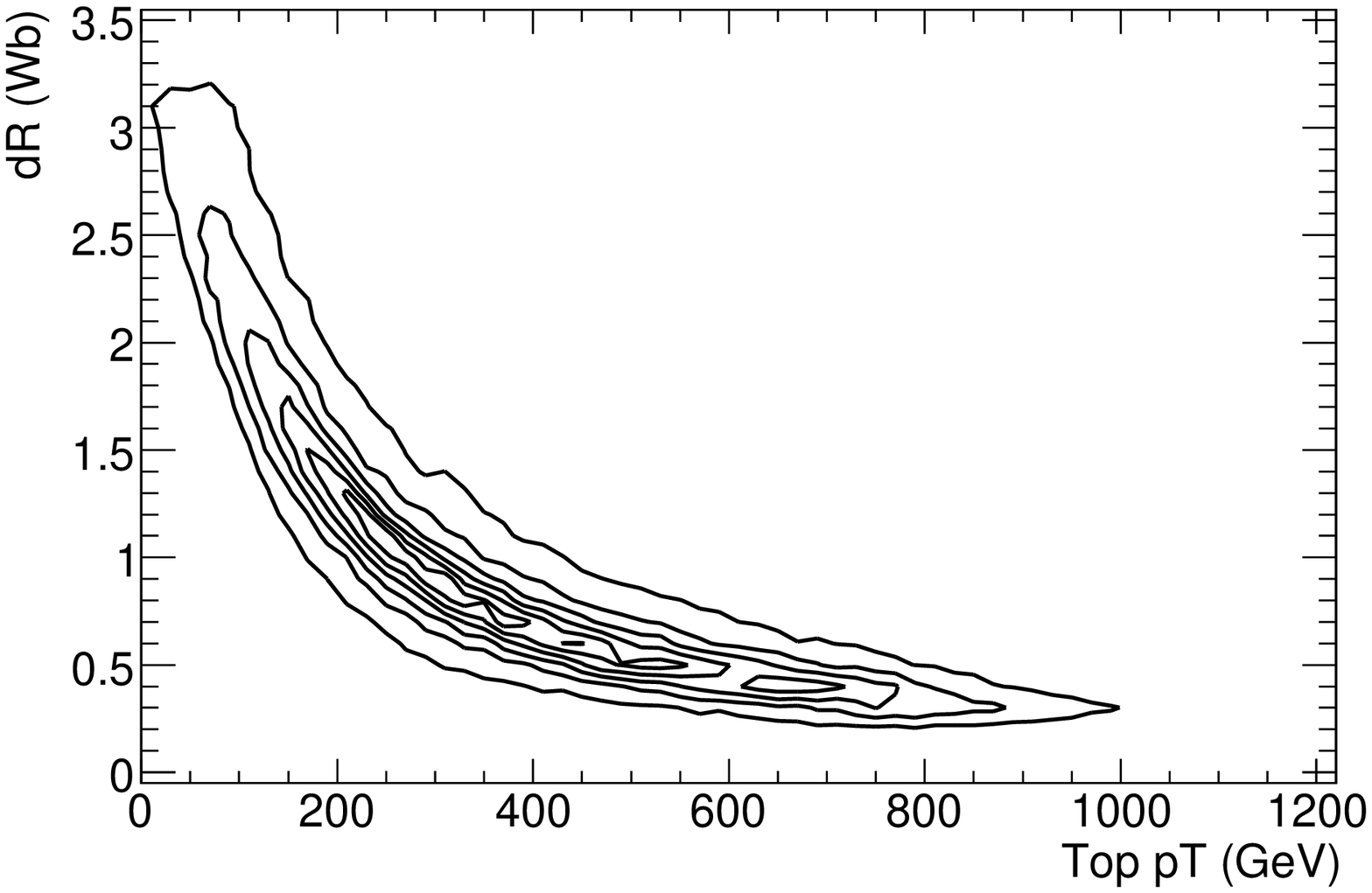}
}
\subfigure[]{
\label{fig:dr2}
\includegraphics[width= 
0.45\textwidth]{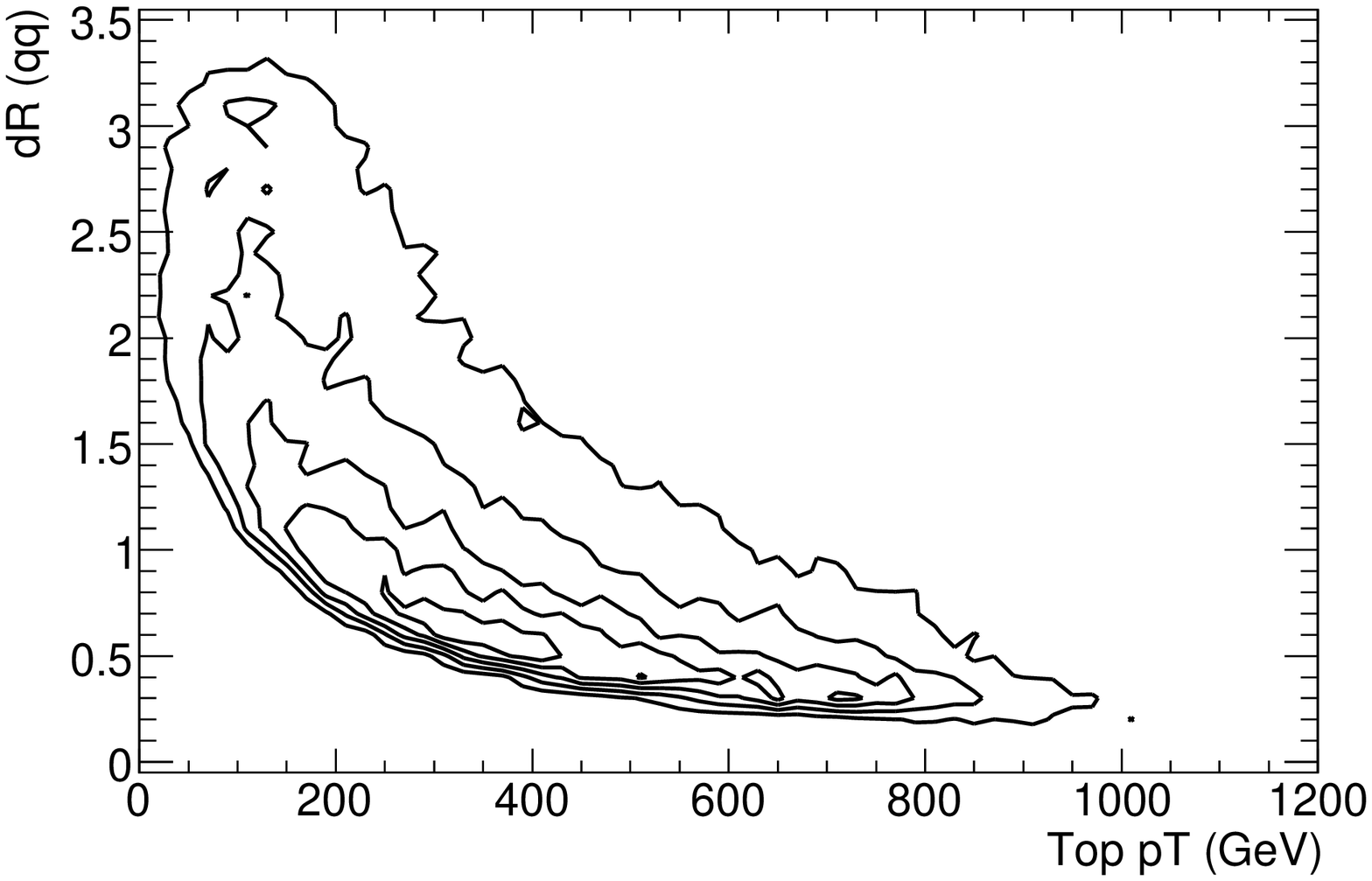}
}
\caption{Angular distances between decay products in top quark decays as a function of
top quark transverse momentum: (a)
between the $b$ quark and $W$ boson, and (b) between quarks from $W$ boson decays.
\label{fig:drbW}}
\end{center}
\end{figure}

The decay of a heavy resonance ($m \gg m_{top}$) to one or more top quarks leads
to a new experimental phenomenology: because of the large top quark momentum, 
its decay products will be collimated~\cite{Lillie:2007yh,Fitzpatrick:2007qr}, 
leading to a merger of the decay products into a single high transverse 
momentum jet, with possibly an embedded charged lepton.  The angular distance 
($\Delta R = \sqrt{(\Delta \phi)^2 + (\Delta \eta)^2}$) between
the decay products as a function of top quark transverse momentum is shown 
in Fig.~\ref{fig:drbW}.  
For top quark transverse momenta larger than about 400 GeV, a large fraction of the 
quarks from hadronic $W$ boson decay are separated by 
$\Delta R <$ 0.5, the typical jet radius.
When the top quark transverse momentum exceeds 600 GeV, the $b$-jet is usually merged 
with the $W$ boson decay products.  New techniques are therefore necessary to 
distinguish jets from high transverse momentum top quark decays from those originating
from light quarks or gluons.  

Even though with standard reconstruction techniques we observe a single, hard jet,
it originated from a massive particle decaying to two or three hard partons.  Therefore,
if it were possible to measure each of the partons perfectly, the direct daughter partons
and the originator's invariant mass could be reconstructed.  Of course, quarks hadronize,
leading to ``cross-talk'', and the detectors are not able to resolve all individual hadrons.
Nevertheless, the fine granularity of the LHC experiments' calorimeters 
can be used to try to resolve the objects inside the jets.  Various
techniques have been 
proposed~\cite{Thaler:2008ju,Kaplan:2008ie,Almeida:2008yp,Agashe:2006hk,Skiba:2007fw,Holdom:2007nw,Butterworth:2008iy,Butterworth:2007ke} and tested in fast simulations.  In the following,
results~\cite{Brooijmans:1077731} obtained using the full simulation of the 
ATLAS detector~\cite{:2008zz}
are presented.

\begin{figure}[htbp]
\begin{center}
\includegraphics[width= 
0.60\textwidth]{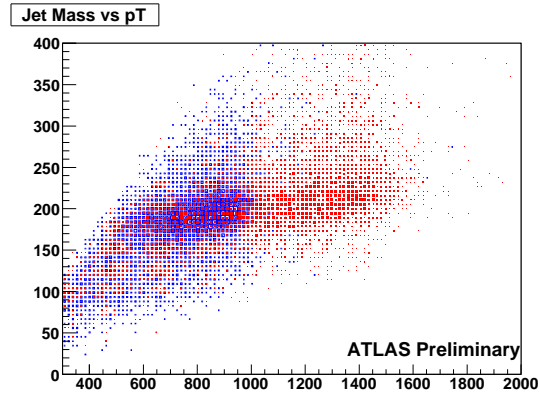}
\caption{Jet mass as a function of transverse momentum for 
``top monojet'': jets from an $m_{Z'} =$ 2 (3) TeV simulated 
$Z'$ sample in blue/open (red/filled).}
\label{fig:jetmvpt}
\end{center}
\end{figure}

A simple discriminating variable is the jet mass, i.e. the invariant mass of all the 
jet constituents.  At the detector level, these constituents are typically calorimeter
cells or protoclusters made from a small number of cells.  (The latter approach allows for 
improved noise suppression and reduces sensitivity to pile-up and underlying event
contributions.)  The jet mass for ``top monojets'', i.e. hadronically decaying top quarks 
in which all decay products are reconstructed as a single jet, is shown in Fig.~\ref{fig:jetmvpt}
as a function of jet transverse momentum.  This plot is based on simulated $Z'$ bosons
with masses $m_{Z'} = $ 2 and 3 TeV.  Events cluster in a band going from $m_{jet} \approx$ 180 GeV for 
jets of 600 GeV transverse momentum, to $m{jet} \approx$ 200 GeV at 1400 GeV. 
There is a slow increase in the jet mass as a function of transverse momentum
due to increased radiation~\footnote{New techniques~\cite{Butterworth:2008iy,Kaplan:2008ie} 
which have been proposed to reduce 
this dependency are being studied in fully simulated events.}, 
but the discrimination power is large since for a given transverse momentum
jets from light quarks or 
gluons have a mass following a negative exponential distribution.  Jet mass, however, 
is insensitive to a potential anisotropy in the jet energy distribution, as 
would be expected when jets from a few hard partons are merged together.

\begin{figure}[htbp]
\begin{center}
\subfigure[]{
\includegraphics[width= 
0.45\textwidth]{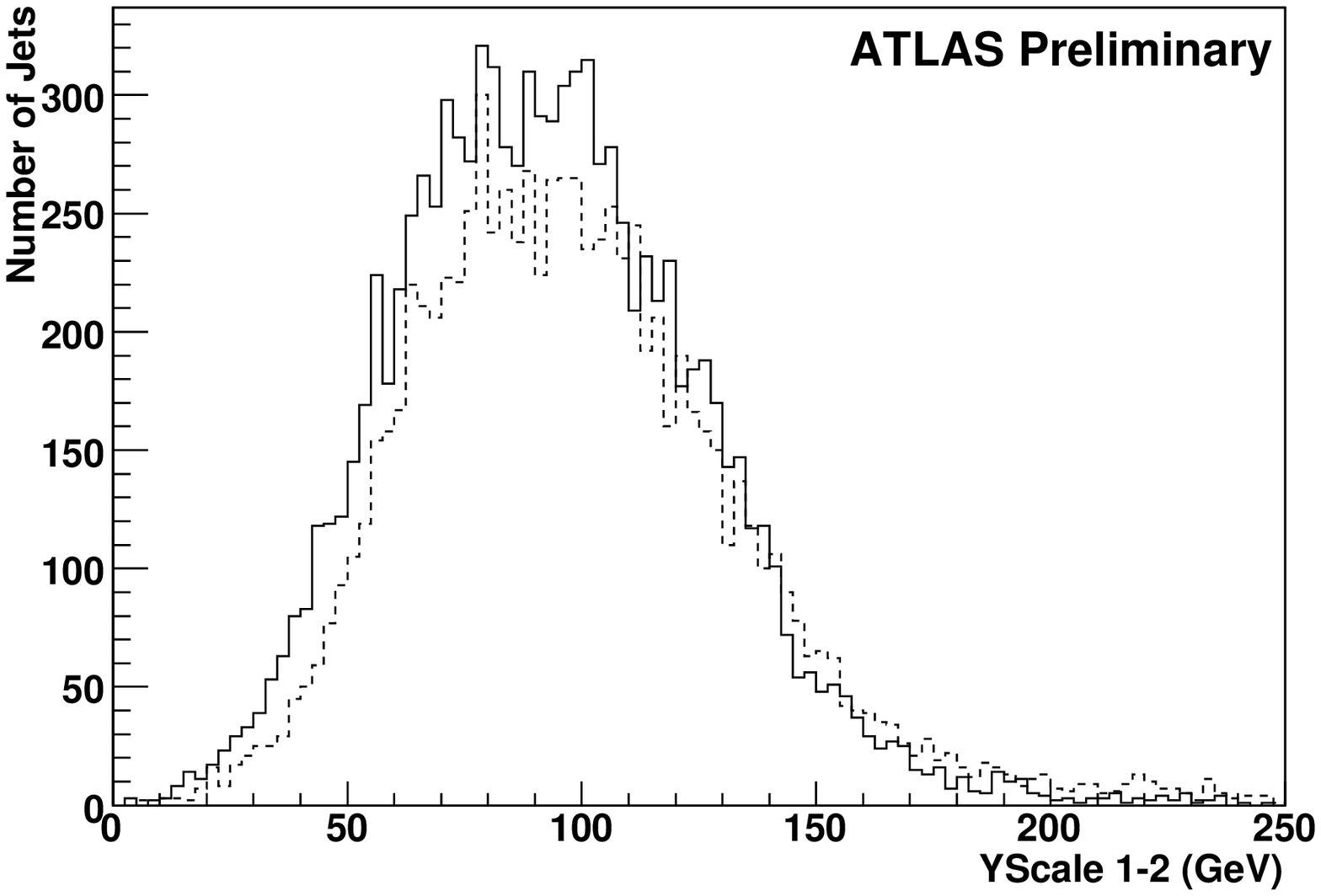}
}
\subfigure[]{
\includegraphics[width=0.45\textwidth]{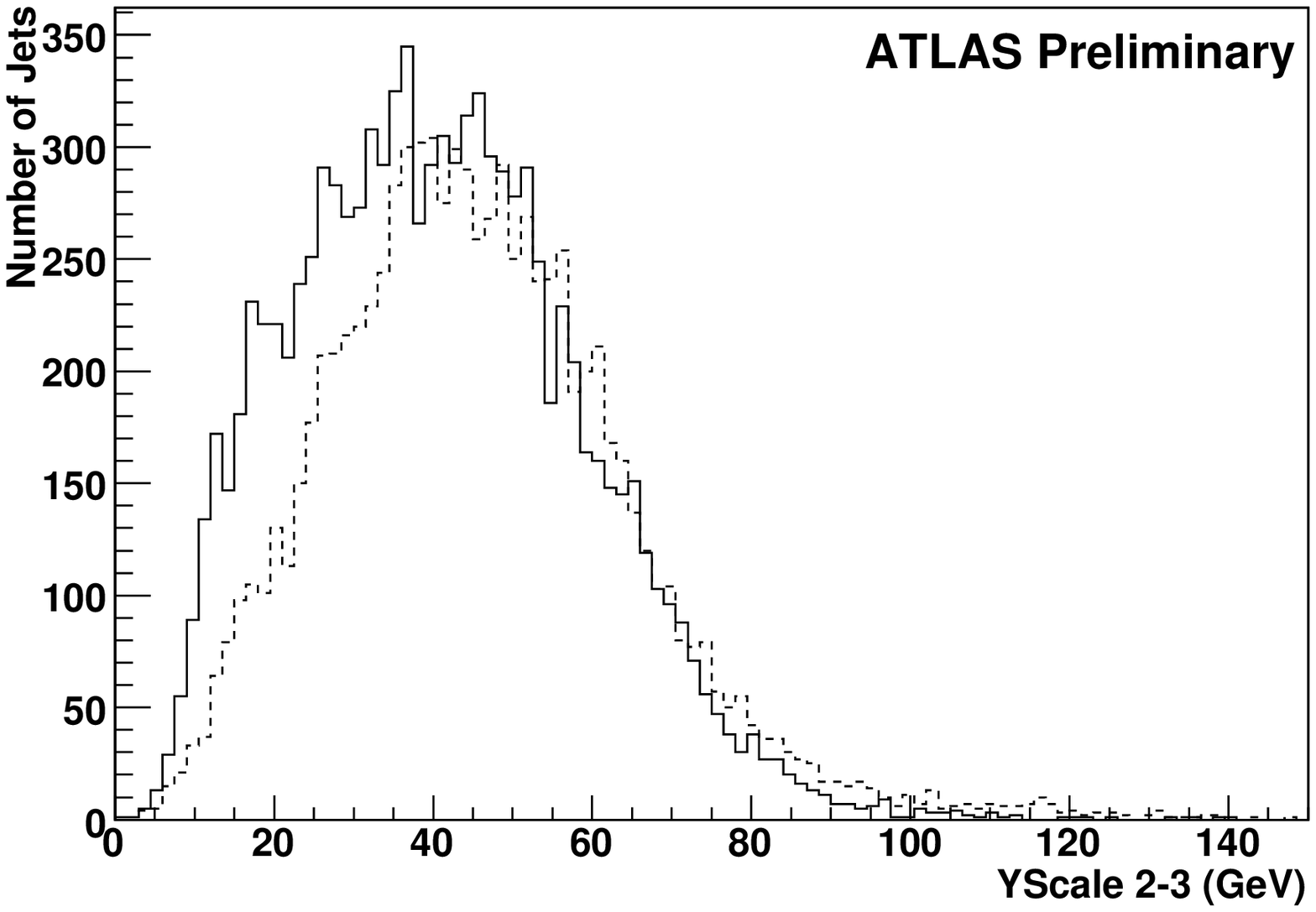}
}
\caption{Scales at which top monojets split into (a) two, and
(b) three jets.  Jets from the $m_{Z'} =$ 2 (3) TeV $Z'$ boson samples are drawn as a solid
(dashed) line.}
\label{fig:ystop}
\end{center}
\end{figure}

$k_\perp$-type jet reconstruction algorithms are well suited to identify substructure in a jet.  As
opposed to the cone-type algorithms which seek to maximize energy in a cone in 
$(\eta,\phi)$-space, $k_\perp$ algorithms are ``nearest neighbor'' clusterers: they 
combine nearby clusters into a jet if their energy-weighted distance is smaller
than a certain quantity.  If a $k_\perp$ jet is formed from mutiple decay products of a heavy particle,
the energy scale at which it splits from one into two (and two into three, etc.) jets,
sometimes called ``$Y-scale$'', is 
related to the mass of that particle.  Figure~\ref{fig:ystop} shows the scales at 
which top monojets split into two or three jets, for both the $m_{Z'}=$ 2 and 3 TeV 
simulated $Z'$ boson samples.  Again, the distributions are very similar, showing
the value of these variables in identifying  high transverse momentum top quarks
in a wide range of momenta.
The scale for splitting into two jets is close to half the top quark mass, and 
for the next splitting it is close to half the $W$ boson mass, as expected.  
For light quark and gluon jets, these distributions are all negative exponentials.

\begin{figure}[htbp]
\begin{center}
\subfigure[]{
\includegraphics[width= 
0.45\textwidth]{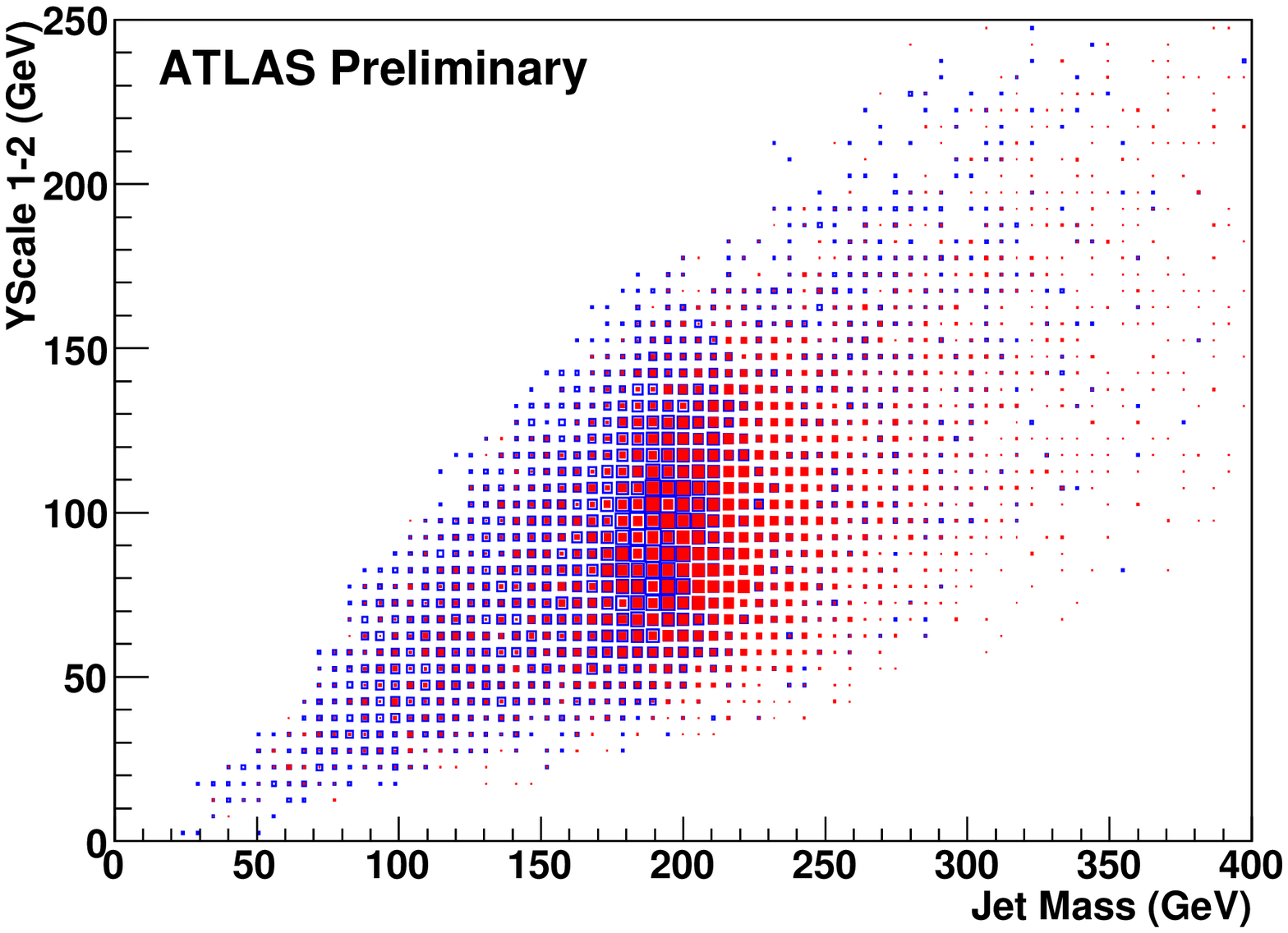}
}
\subfigure[]{
\includegraphics[width= 
0.45\textwidth]{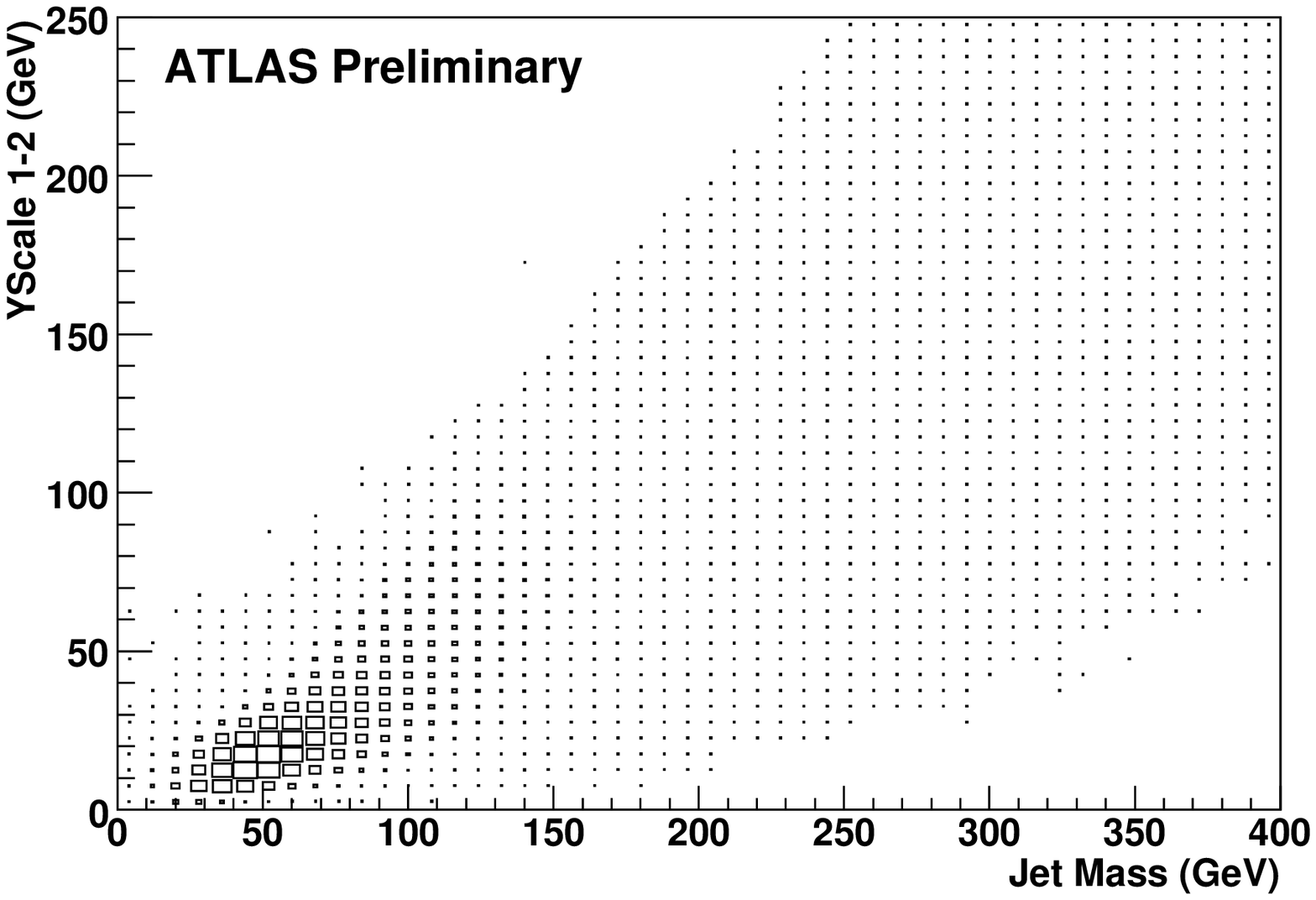}
}
\caption{Correlation between $Y-scale$ values and jet mass for splitting into two jets
for (a) top monojets and (b) light quark and gluon jets.} 
\label{fig:ysvjm-top}
\end{center}
\end{figure}

\begin{figure}[htbp]
\begin{center}
\includegraphics[width= 
0.6\textwidth]{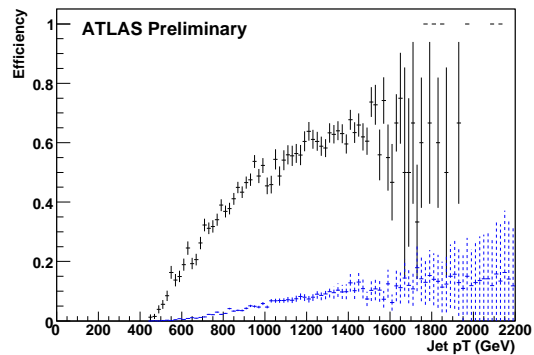}
\caption{Selection efficiency for top monojets (solid, black) and light quark and gluon 
jets in the (blue, dashed) as a function of reconstructed jet transverse momentum.}
\label{fig:effs}
\end{center}
\end{figure}

The correlation between jet mass and splittings into two, three and four jets can 
be exploited to discriminate between top and light quark and gluon jets.  This is 
shown for splitting in two jets and jet mass in Fig.~\ref{fig:ysvjm-top}.
The efficiencies resulting from simple two-dimensional cuts in these variables 
for signal and background are shown in Fig.~\ref{fig:effs}.  Further studies are 
underway and an improvement in signal/background of a factor of three or more is 
expected.  In the lepton-plus-jets channel, this will make QCD production
of $t\bar{t}$ pairs the dominant background in the search for heavy resonances decaying
to top quarks pairs.  $t\bar{t}$ invariant mass resolution then becomes the most 
critical aspect in isolating a signal.

\section{Final Remarks}

Many models of new physics predict the existence of resonances which will be detectable 
at the LHC.  The sensitivity of the experiments to a number of these has been exposed
in this paper.  Significant information on the underlying physics will be available 
immediately through the observed production cross-section and decay mode(s) of the 
dicovery signal, but  
this will need to be followed up by searches in complementary channels and detailed
measurements of signal properties.  Hopefully, this will allow us to learn something 
about the origins of the properties of fermions, in addition to opening a window on 
physics at the TeV scale and beyond.

\section*{Acknowledgments}

The author would like to thank the authors of all the studies presented in this review.
Understanding of the capabilities of the LHC detectors and the corresponding 
discovery reach is not only important for model builders, but also a crucial part 
in preparing for the real data analysis.  This review will hopefully be useful 
for both newcomers to the field as well as more advanced particle physicists.


\bibliographystyle{ws-mpla}
\bibliography{procs}

\begin{thebibliography}{10}

\bibitem{Alcaraz:2007ri}
J.~Alcaraz {\em et~al.}, {\em eprint} {\bf hep-ex/0712.0929} (2007).

\bibitem{:2005ema}
{The LEP and SLD Collaborations}, {\em Phys. Rept.} {\bf 427}, p. 257 (2006).

\bibitem{Lee:1977eg}
B.~W. Lee, C.~Quigg and H.~B. Thacker, {\em Phys. Rev.} {\bf D16}, p. 1519
  (1977).

\bibitem{Higgs:1964ia}
P.~W. Higgs, {\em Phys. Lett.} {\bf 12}, 132 (1964).

\bibitem{Higgs:1964pj}
P.~W. Higgs, {\em Phys. Rev. Lett.} {\bf 13}, 508 (1964).

\bibitem{Higgs:1966ev}
P.~W. Higgs, {\em Phys. Rev.} {\bf 145}, 1156 (1966).

\bibitem{Englert:1964et}
F.~Englert and R.~Brout, {\em Phys. Rev. Lett.} {\bf 13}, 321 (1964).

\bibitem{Guralnik:1964eu}
G.~S. Guralnik, C.~R. Hagen and T.~W.~B. Kibble, {\em Phys. Rev. Lett.} {\bf
  13}, 585 (1964).

\bibitem{Shrock:2008sb}
R.~Shrock, {\em eprint} {\bf arXiv:0809.0087} (2008).

\bibitem{Martin:1997ns}
S.~P. Martin, {\em eprint} {\bf hep-ph/9709356} (1997).

\bibitem{ArkaniHamed:2001nc}
N.~Arkani-Hamed, A.~G. Cohen and H.~Georgi, {\em Phys. Lett.} {\bf B513}, 232
  (2001).

\bibitem{ArkaniHamed:1998rs}
N.~Arkani-Hamed, S.~Dimopoulos and G.~R. Dvali, {\em Phys. Lett.} {\bf B429},
  263 (1998).

\bibitem{Randall:1999ee}
L.~Randall and R.~Sundrum, {\em Phys. Rev. Lett.} {\bf 83}, 3370 (1999).

\bibitem{Rizzo:2006nw}
T.~G. Rizzo, {\em eprint} {\bf hep-ph/0610104} (2006).

\bibitem{cdfee08}
{The CDF Collaboration}, {High-Mass Dielectron Resonance Search in $p \bar p$
  Collisions at $\sqrt{s}$ = 1.96 TeV}, CDF/PUB/EXOTIC/PUBLIC/9160, (2008).

\bibitem{cdfmumu08}
{The CDF Collaboration}, {A Search for Dimuon Resonances with CDF in Run II},
  \href{http://www-cdf.fnal.gov/physics/exotic/r2a/20080710.dimuon_resonance/},
  (2008).

\bibitem{atlascsc}
{ATLAS Collaboration}, {\em Expected Performance of the ATLAS Experiment,
  Detector, Trigger and Physics}, Tech. Rep. CERN-OPEN-2008-020, CERN (Geneva,
  2008 (to appear)).

\bibitem{Ball:2007zza}
G.~L. Bayatian {\em et~al.}, {\em J. Phys.} {\bf G34}, 995 (2007).

\bibitem{:2007sb}
T.~Aaltonen {\em et~al.}, {\em Phys. Rev. Lett.} {\bf 99}, p. 171802 (2007).

\bibitem{Allanach:2000nr}
B.~C. Allanach, K.~Odagiri, M.~A. Parker and B.~R. Webber, {\em JHEP} {\bf 09},
  p. 019 (2000).

\bibitem{Cousins:916380}
R.~Cousins, J.~Mumford and V.~Valuev, {\em Measurement of Forward-Backward
  Asymmetry of Simulated and Reconstructed $Z' \to \mu^{+}\mu^{-}$ Events in
  CMS}, Tech. Rep. CMS-NOTE-2005-022. CERN-CMS-NOTE-2005-022, CERN (Geneva,
  2005).

\bibitem{Henriques:682130}
A.~Henriques and L.~Poggioli, {\em Detection of the $Z'$ vector boson in the
  jet decay mode $(Z' \to q \bar{q}) (g) \to jj$. Resolution and pile-up
  studies}, Tech. Rep. ATL-PHYS-92-010. ATL-GE-PN-10, CERN (Geneva, 1992).

\bibitem{Ferrari:684146}
A.~Ferrari and J.~Collot, {\em Sensitivity study for a Z' boson decaying into
  two right-handed Majorana neutrinos at LHC in the ATLAS detector}, Tech. Rep.
  ATL-PHYS-2000-034, CERN (Geneva, 2000).

\bibitem{Rizzo:2007xs}
T.~G. Rizzo, {\em JHEP} {\bf 05}, p. 037 (2007).

\bibitem{Arik:684195}
E.~Arik, O.~Celik, A.~Mailov, D.~Benchekroun, A.~Hoummada, J.~Collot and
  A.~Ferrari, {\em A study of $pp \to W' \to WZ$ at LHC in the ATLAS
  experiment}, Tech. Rep. ATL-PHYS-2001-005, CERN (Geneva, 2001).

\bibitem{GonzálezdelaHoz:814346}
S.~Gonz\'alez de~la Hoz, L.~March and E.~Ros, {\em Search for hadronic decays
  of $Z_{H}$ and $W_{H}$ in the Little Higgs model}, Tech. Rep.
  ATL-PHYS-PUB-2006-003. ATL-COM-PHYS-2005-001, CERN (Geneva, 2005).

\bibitem{Langacker:2008yv}
P.~Langacker, {\em eprint} {\bf 0801.1345} (2008).

\bibitem{Mehdiyev:2006tz}
R.~Mehdiyev, S.~Sultansoy, G.~Unel and M.~Yilmaz, {\em Eur. Phys. J.} {\bf
  C49}, 613 (2007).

\bibitem{Mehdiyev:2007pf}
R.~Mehdiyev, A.~Siodmok, S.~Sultansoy and G.~Unel, {\em Eur. Phys. J.} {\bf
  C54}, 507 (2008).

\bibitem{Abazov:2008kp}
V.~M. Abazov {\em et~al.}, {\em Phys. Rev. Lett.} {\bf 101}, p. 011601 (2008).

\bibitem{Aaltonen:2008hh}
T.~Aaltonen {\em et~al.}, {\em eprint} {\bf 0807.3132} (2008).

\bibitem{Abazov:2008as}
V.~M. Abazov {\em et~al.}, {\em eprint} {\bf 0809.2813} (2008).

\bibitem{Abazov:2007ra}
V.~M. Abazov {\em et~al.}, {\em Phys. Rev. Lett.} {\bf 100}, p. 091802 (2008).

\bibitem{Davoudiasl:2000wi}
H.~Davoudiasl, J.~L. Hewett and T.~G. Rizzo, {\em Phys. Rev.} {\bf D63}, p.
  075004 (2001).

\bibitem{Agashe:2003zs}
K.~Agashe, A.~Delgado, M.~J. May and R.~Sundrum, {\em JHEP} {\bf 08}, p. 050
  (2003).

\bibitem{Lillie:2007yh}
B.~Lillie, L.~Randall and L.-T. Wang, {\em JHEP} {\bf 09}, p. 074 (2007).

\bibitem{Djouadi:2007eg}
A.~Djouadi, G.~Moreau and R.~K. Singh, {\em Nucl. Phys.} {\bf B797}, 1 (2008).

\bibitem{Fitzpatrick:2007qr}
A.~L. Fitzpatrick, J.~Kaplan, L.~Randall and L.-T. Wang, {\em JHEP} {\bf 09},
  p. 013 (2007).

\bibitem{Thaler:2008ju}
J.~Thaler and L.-T. Wang, {\em JHEP} {\bf 07}, p. 092 (2008).

\bibitem{Kaplan:2008ie}
D.~E. Kaplan, K.~Rehermann, M.~D. Schwartz and B.~Tweedie, {\em eprint} {\bf
  0806.0848} (2008).

\bibitem{Almeida:2008yp}
L.~G. Almeida {\em et~al.}, {\em eprint} {\bf 0807.0234} (2008).

\bibitem{Agashe:2006hk}
K.~Agashe, A.~Belyaev, T.~Krupovnickas, G.~Perez and J.~Virzi, {\em Phys. Rev.}
  {\bf D77}, p. 015003 (2008).

\bibitem{Skiba:2007fw}
W.~Skiba and D.~Tucker-Smith, {\em Phys. Rev.} {\bf D75}, p. 115010 (2007).

\bibitem{Holdom:2007nw}
B.~Holdom, {\em JHEP} {\bf 03}, p. 063 (2007).

\bibitem{Butterworth:2008iy}
J.~M. Butterworth, A.~R. Davison, M.~Rubin and G.~P. Salam, {\em Phys. Rev.
  Lett.} {\bf 100}, p. 242001 (2008).

\bibitem{Butterworth:2007ke}
J.~M. Butterworth, J.~R. Ellis and A.~R. Raklev, {\em JHEP} {\bf 05}, p. 033
  (2007).

\bibitem{Brooijmans:1077731}
G.~Brooijmans, {\em High pT Hadronic Top Quark Identification}, Tech. Rep.
  ATL-PHYS-CONF-2008-008. ATL-COM-PHYS-2008-001, CERN (Geneva, 2008).

\bibitem{:2008zz}
S.~Bentvelsen {\em et~al.}, {\em JINST} {\bf 3}, p. S08003 (2008).

\end{thebibliography}

\end{document}